\documentclass[journal]{IEEEtran}
\usepackage{cite}
\usepackage{amsmath,amssymb,amsfonts}
\usepackage{algorithmic}
\usepackage{graphicx}
\usepackage{booktabs}
\usepackage{textcomp}
\usepackage{xcolor}
\usepackage{url}
\usepackage{cleveref}
\usepackage{xcolor}
\usepackage{listings}
\usepackage[breakable]{tcolorbox}

\newcommand{\approach}{\textsc{SPDDwL}}

\newtcolorbox{rqbox}{breakable,left=4pt,right=4pt,top=4pt,bottom=4pt}
\newcommand\code[1]{{\tt\small #1}}
\definecolor{dkgreen}{rgb}{0,0.3,0}
\definecolor{cmtgreen}{rgb}{0,0.6,0}
\definecolor{ltblue}{rgb}{0,0.4,0.4}
\definecolor{dkviolet}{rgb}{0.3,0,0.5}
\definecolor{dkblue}{rgb}{0,0.2,0.2}
\definecolor{dkred}{rgb}{0.6,0,0}
\lstdefinelanguage{Coq}{ 
    mathescape=true,
    texcl=false, 
    escapeinside={(@}{@)},
    morekeywords=[1]{Section, Module, End, Require, Import, Export,
        Variable, Variables, Parameter, Parameters, Axiom, Hypothesis,
        Hypotheses, Notation, Local, Tactic, Reserved, Scope, Open, Close,
        Bind, Delimit, Definition, Let, Ltac, Fixpoint, CoFixpoint, Add,
        Morphism, Relation, Implicit, Arguments, Unset, Contextual,
        Strict, Prenex, Implicits, Inductive, CoInductive, Record,
        Structure, Canonical, Coercion, Context, Class, Global, Instance,
        Program, Infix, Theorem, Lemma, Corollary, Proposition, Fact,
        Remark, Example, Proof, Goal, Save, Qed, Defined, Hint, Resolve,
        Rewrite, View, Search, Show, Print, Printing, All, Eval, Check,
        Projections, inside, outside, Def},
    morekeywords=[2]{forall, exists, exists2, fun, fix, cofix, struct,
        match, with, end, as, in, return, let, if, is, then, else, for, of,
        nosimpl, when},
    morekeywords=[3]{Type, Prop, Set, true, false, option},
    morekeywords=[4]{pose, set, move, case, elim, apply, clear, hnf,
        intro, intros, generalize, rename, pattern, after, destruct,
        induction, using, refine, inversion, injection, rewrite, congr,
        unlock, compute, ring, field, fourier, replace, fold, unfold,
        change, cutrewrite, simpl, have, suff, wlog, suffices, without,
        loss, nat_norm, assert, cut, trivial, revert, bool_congr, nat_congr,
        symmetry, transitivity, auto, split, left, right, autorewrite},
    morekeywords=[5]{by, done, exact, reflexivity, tauto, romega, omega,
        assumption, solve, contradiction, discriminate},
    morekeywords=[6]{do, last, first, try, idtac, repeat},
    morecomment=[s]{(*}{*)},
    showstringspaces=false,
    morestring=[b]",
    morestring=[d]’,
    tabsize=3,
    extendedchars=false,
    sensitive=true,
    breaklines=false,
    basicstyle=\fontsize{9.5pt}{11.4pt}\selectfont\ttfamily,
    captionpos=b,
    columns=[l]flexible,
    identifierstyle={\ttfamily\color{black}},
    keywordstyle=[1]{\bfseries\ttfamily\color{dkviolet}},
    keywordstyle=[2]{\bfseries\ttfamily\color{dkgreen}},
    keywordstyle=[3]{\bfseries\ttfamily\color{ltblue}},
    keywordstyle=[4]{\bfseries\ttfamily\color{dkblue}},
    keywordstyle=[5]{\bfseries\ttfamily\color{dkred}},
    stringstyle=\ttfamily,
    commentstyle={\bfseries\ttfamily\color{cmtgreen}},
    literate=
    {\\forall}{{\color{dkgreen}{$\forall\;$}}}1
    {\\exists}{{$\exists\;$}}1
    {<-}{{$\leftarrow\;$}}1
    {=>}{{$\Rightarrow\;$}}1
    {==}{{\code{==}\;}}1
    {==>}{{\code{==>}\;}}1
    {->}{{$\rightarrow\;$}}1
    {<->}{{$\leftrightarrow\;$}}1
    {<==}{{$\leq\;$}}1
    {\#}{{$^\star$}}1 
    {\\o}{{$\circ\;$}}1 
    {\@}{{$\cdot$}}1 
    {\/\\}{{$\wedge\;$}}1
    {\\\/}{{$\vee\;$}}1
    {++}{{\code{++}}}1
    {~}{{$\sim$}}1
    {\@\@}{{$@$}}1
    {\\mapsto}{{$\mapsto\;$}}1
    {\\hline}{{\rule{\linewidth}{0.5pt}}}1
}[keywords,comments,strings]

\lstdefinelanguage{PlainText}{ 
    mathescape=true,
    texcl=false, 
    escapeinside={(@}{@)},
    morecomment=[s]{(*}{*)},
    showstringspaces=false,
    morestring=[b]",
    morestring=[d]’,
    tabsize=3,
    extendedchars=false,
    sensitive=true,
    breaklines=false,
    basicstyle=\fontsize{5.5pt}{7.5pt}\selectfont\ttfamily,
    captionpos=b,
    columns=[l]flexible,
    identifierstyle={\ttfamily\color{black}},
    keywordstyle=[1]{\bfseries\ttfamily\color{dkviolet}},
    keywordstyle=[2]{\bfseries\ttfamily\color{dkgreen}},
    keywordstyle=[3]{\bfseries\ttfamily\color{ltblue}},
    keywordstyle=[4]{\bfseries\ttfamily\color{dkblue}},
    keywordstyle=[5]{\bfseries\ttfamily\color{dkred}},
    stringstyle=\ttfamily,
    commentstyle={\bfseries\ttfamily\color{dkgreen}},
    literate=
    {\\forall}{{\color{dkgreen}{$\forall\;$}}}1
    {\\exists}{{$\exists\;$}}1
    {<-}{{$\leftarrow\;$}}1
    {=>}{{$\Rightarrow\;$}}1
    {==}{{\code{==}\;}}1
    {==>}{{\code{==>}\;}}1
    {->}{{$\rightarrow\;$}}1
    {<->}{{$\leftrightarrow\;$}}1
    {<==}{{$\leq\;$}}1
    {\#}{{$^\star$}}1 
    {\\o}{{$\circ\;$}}1 
    {\@}{{$\cdot$}}1 
    {\/\\}{{$\wedge\;$}}1
    {\\\/}{{$\vee\;$}}1
    {++}{{\code{++}}}1
    {~}{{$\sim$}}1
    {\@\@}{{$@$}}1
    {\\mapsto}{{$\mapsto\;$}}1
    {\\hline}{{\rule{\linewidth}{0.5pt}}}1
}[keywords,comments,strings]

\ifCLASSINFOpdf
\else
\fi
\begin{document}
%
\title{Trustworthy Software Project Generation : a Case Study with an Interactive Theorem Prover}
%
%
%

\author{%
  \begin{minipage}[t]{0.45\linewidth}
    \centering
    {\large\bfseries Jian Fang}\\[4pt]
    \textit{Key Laboratory of High Confidence Software Technologies (Peking University), Ministry of Education; School of Computer Science}\\
    \textit{Peking University}\\
    Beijing, China\\
    \texttt{fangjian@stu.pku.edu.cn}
  \end{minipage}%
  \hfill
  \begin{minipage}[t]{0.45\linewidth}
    \centering
    {\large\bfseries Yingfei Xiong}\\[4pt]
    \textit{Key Laboratory of High Confidence Software Technologies (Peking University), Ministry of Education; School of Computer Science}\\
    \textit{Peking University}\\
    Beijing, China\\
    \texttt{xiongyf@pku.edu.cn}
  \end{minipage}%
}


%
%

\markboth{Journal of \LaTeX\ Class Files,~Vol.~14, No.~8, August~2015}%
{Shell \MakeLowercase{\textit{et al.}}: Bare Demo of IEEEtran.cls for IEEE Journals}
%



\maketitle

\begin{abstract}
Generating code from natural-language requirements has become a primary route
for LLM-assisted software development. Although LLMs can successfully complete
small programming tasks, generating an entire complex project remains
unreliable because subtle errors may survive compilation and testing.
Verified programming can reduce this risk by requiring generated
implementations to satisfy machine-checked specifications. Existing
explorations mostly target verification-oriented languages and toolchains
such as Dafny and Frama-C, but directly generating project-scale verified
code with these systems remains difficult.

This paper investigates whether the programming facilities of interactive
theorem provers (ITPs) provide a viable alternative for developing large
software projects. ITPs can define and prove pure total functions, but do not
directly support common effects such as I/O. Our agent therefore first derives
a coding plan that separates effectful components from as much pure logic as
possible. Effectful modules are implemented in the target language; for pure
modules, the agent generates formal specifications, implementations in the
functional language of an ITP, and machine-checked proofs, then extracts
ordinary target-language code for integration into the final software.

We study this route through the fully automatic development of a CPU
interpreter for all 47 instructions of the unprivileged RISC-V RV32I base:
after the requirements are supplied, no human intervenes in
synthesis, proof repair, extraction, or integration. With Rocq as the
backend, the agent completes the project within 30 minutes, producing 1,859
lines of verified Rocq and extracting 2,848 lines of C++. The resulting
interpreter passes all 265 LLM-generated tests covering the 47 instructions
and exhibits zero crashes and zero hangs during 12 hours of AFL++ fuzzing.

To the best of our knowledge, this is the largest reported realistic software
project with a machine-checked verified core developed fully automatically by
an LLM agent. Under the same 30-minute configuration, a Dafny-based backend
fails to complete verification.
Our observation is that Rocq exposes a concrete proof state when a proof
attempt fails, giving the agent actionable feedback for repair. These results
provide empirical evidence that ITP-based verified programming is a feasible
route for LLM-generated software projects.
\end{abstract}


\section{Introduction}

Generating code from natural-language requirements has become a primary
route for LLM-assisted software development, supported by code models and
coding agents that can edit, execute, and test software
projects~\cite{guo2024deepseekcoderlargelanguagemodel,chen2021evaluatinglargelanguagemodels,claude_code_overview,cursor_agents}.
This route is effective for many small tasks, but generating a complete
complex project remains difficult to trust. LLM-generated code can compile,
type-check, and pass the available tests while still violating intended
semantics~\cite{li2026fmagent,DBLP:journals/tois/HuangYMZFWCPFQL25}.
The problem is particularly acute when generated software implements
critical state transitions or other behaviors for which missed corner cases
are costly.

Verified programming provides a possible response: instead of accepting code
after testing alone, require generated implementations to satisfy formal,
machine-checked specifications. Existing LLM-assisted work has explored
formal constraints and generated annotations for automated verification
systems~\cite{DBLP:journals/pacmpl/MundlerHWSSV25,beg2026evaluatingllmgeneratedacslannotations}.
Verification-oriented languages such as Dafny~\cite{DBLP:conf/lpar/Leino10}
are natural targets for this direction, but directly producing verified code
for a complete project remains demanding for an LLM agent, especially when a
failed automated-verification attempt provides little local information for
repair.

Interactive theorem provers (ITPs), such as Rocq~\cite{CoqRefMan900}, Lean~\cite{DBLP:conf/cade/Moura021}, and Isabelle~\cite{DBLP:conf/tphol/WenzelPN08},
provide another route.
In addition to proving mathematical theorems, an ITP can define executable
functional programs, state properties about those programs, check explicit
proofs, and, for suitable backends, extract verified definitions into
ordinary development languages. This raises the question studied in this
paper: \emph{can an LLM agent use the programming facilities of an ITP to
develop a complete, non-trivial software project with verified core
functionality?}

There is a practical obstacle. ITP programming languages are primarily
designed for pure functions and proofs, while executable projects also
require I/O, allocation, interaction with the runtime, and other effects.
We therefore explore an architecture that maximizes the pure portion of the
project: an agent derives a coding plan that separates effectful host
components from pure core logic; it specifies, implements, and proves the
pure components in an ITP; it then extracts those components and integrates
them with target-language code that provides the effects. The formal
guarantee applies to the extracted components and the properties proved
about them, while the integrated executable is additionally validated by
testing and fuzzing.

We implement this workflow as \approach{} and instantiate it with Rocq for
our primary study. In the Rocq instantiation, the workflow proceeds as
follows:
\begin{itemize}
    \item A requirement-analysis agent turns natural-language requirements
    into a coding plan, identifying pure functions, host-side effects, and
    the formal properties of the pure functions.
    \item A coding agent generates Rocq functional definitions for the
    verified core, and a proving agent generates Rocq propositions and tactic
    proofs for the plan's properties.
    \item Rocq checks the definitions and proofs. When checking fails, the
    proof state and diagnostic feedback are returned to the responsible
    agent for repair.
    \item Accepted definitions are extracted to the target language and
    linked with a small host layer implementing the effectful operations.
\end{itemize}
The architecture is not specific to Rocq: it requires an ITP with an
executable functional language, machine-checked proofs, and a path for
deploying the verified definitions. Rocq is the backend through which we
evaluate the route concretely.

\noindent\textbf{Evaluation.}
We ask \approach{} to construct a CPU interpreter for the 47
instructions of the unprivileged RISC-V RV32I base. With Claude Opus~4.7
and Rocq~9.0.1 as the backend, the agent completes development within 30
minutes. This run is fully automatic: once we supply the requirements and
target configuration, no human intervenes in the generated
definitions, specifications, proofs or repairs, extraction, or C++ host
integration. The agent synthesizes 1,859 lines of verified Rocq and extracts
2,848 lines of C++. All 265 LLM-generated instruction tests pass on the
extracted interpreter; a 12-hour AFL++ campaign executes 98.2 million inputs
and finds zero crashes and zero hangs. Closest related studies generate
collections of separate verified programs or verified algorithmic
tasks~\cite{bursuc2025vericoding,baksys2026atlas,gloeckle2026wybecoder}, or
develop a substantial proof artifact with human
guidance~\cite{paraskevopoulou2026machinegeneratedmachinecheckedproofsverified}.
To the best of our knowledge, measured by the machine-checked core, this is
the largest reported runnable software project with a verified core developed
fully automatically by an LLM agent. In a matched 30-minute run with a Dafny
backend, the agent does not complete verification. Inspection of the failed
run indicates that Rocq's explicit proof states give the agent a more
actionable repair signal than the observed solver timeouts in the Dafny run.

\noindent\textbf{Contributions.}
This paper makes the following contributions:
\begin{itemize}
    \item We provide what is, to the best of our knowledge, the largest
    reported runnable software project with a machine-checked core developed
    fully automatically by an LLM agent: a RISC-V CPU interpreter
    whose functional core is generated, proved, extracted, and integrated
    without human intervention.
    \item We articulate and instantiate a workflow that separates pure,
    formally specified components from effectful host components, enabling
    proof and extraction to be used within a runnable software project.
    \item We compare Rocq and Dafny backends under the same synthesis task
    and time budget. The result identifies concrete proof-state feedback as
    a promising property of ITP backends for agent-driven development,
    without claiming that one verifier dominates in all settings.
\end{itemize}

\section{Background}

\noindent\textbf{Interactive theorem provers as programming backends.}
Interactive theorem provers (ITPs), including Rocq, Lean, and
Isabelle/HOL~\cite{CoqRefMan900,DBLP:conf/cade/Moura021,DBLP:conf/tphol/WenzelPN08},
let developers define executable functions and prove logical properties
about those definitions. Unlike a test suite, which exercises finitely many
inputs, a checked proof establishes its stated property for every input
covered by the theorem. ITPs are interactive because proof construction
exposes a goal state: after each proof step, the developer sees the
remaining obligations. For an LLM agent, this state can also act as
structured feedback for subsequent repair attempts.

\noindent\textbf{Rocq in this study.}
Our evaluated ITP backend is Rocq~\cite{CoqRefMan900} (formerly Coq), which
is based on the Calculus of Inductive Constructions
(CIC)~\cite{paulin2015introduction}. Its language serves both as a typed
functional programming language, with inductive datatypes and recursive
definitions, and as a logic in which propositions are types and proofs are
terms. A program property in Rocq may describe a computed value, a state
invariant, or a frame condition asserting what a transformation leaves
unchanged. The Rocq kernel accepts that property only after all proof goals
have been discharged.

\noindent\textbf{Side Effect and Pure Definitions.}
Definitions directly reasoned about in Rocq are pure. A deployed
software project, however, also performs operations such as file or console
I/O, allocation, and communication with its runtime environment. Although
such effects can be represented inside proof assistants using semantic
models such as Interaction Trees~\cite{10.1145/3371119}, this paper studies
a simpler project architecture: behavior that can be expressed as
deterministic transformations of explicit data is placed in the verified
functional core, whereas concrete effects are implemented by a target-language
host layer. For the CPU interpreter, instruction decoding and architectural
state transformations belong to the pure core; the C++ layer supplies
integration and effectful services. Consequently, machine-checked
guarantees cover the extracted core relative to its specifications, not every
line of the complete executable.

\noindent\textbf{Code Extraction.}
After verification, computational definitions can be extracted to an
ordinary target language while proof terms are erased. In the Rocq
instantiation, extraction translates the verified definitions into C++ for
integration with the host program. This use of extraction follows the same
general principle employed in verified systems such as
CompCert~\cite{Leroy-BKSPF-2016} and CertiCoq~\cite{anand2017certicoq}.
Our evaluation assumes that the extraction backend preserves the behavior of
the accepted Rocq definitions; dynamic validation evaluates the integrated
program, including its unverified host boundary.

\section{Overview}\label{sec:overview}

\begin{figure*}[t]
\center
\includegraphics[width=\textwidth]{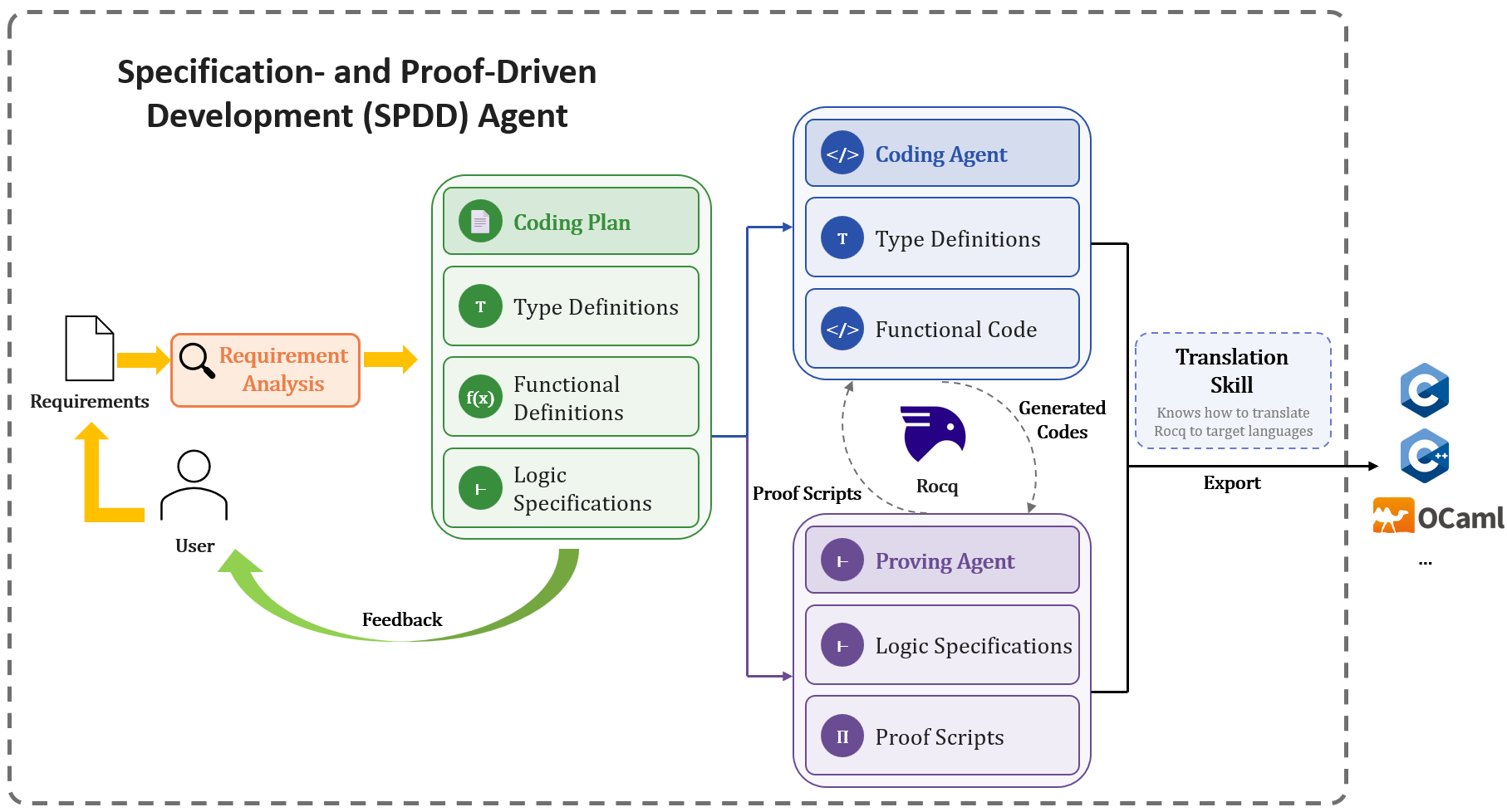}
\caption{The \approach{} workflow instantiated with Rocq.}
\label{fig:overview}
\end{figure*}

\approach{} is an ITP-based workflow for turning natural-language
requirements into a verified functional core that can be linked into a
software project. It is parameterized by an ITP backend and an extraction
path to the target development language. Our implementation uses
Rocq~\cite{CoqRefMan900} and extracts C++ code~\cite{bloomberg_crane_maintainers}; accordingly, the concrete
running example in this section is written in Rocq. The coding agent and
proving agent synthesize an implementation, its specification, and its
proof, and the extracted definitions are exposed to the host project through
APIs. The overall workflow is shown in Figure~\ref{fig:overview}.

CPU emulators of this kind are widely used in simulation and
firmware development~\cite{bartholomew2006qemu,quynh2015unicorn}.
Consider the requirement in the~\Cref{fig:jalr-req}, which
describes a single instruction taken from a larger project that
implements a RISC-V emulator~\cite{riscv_unpriv_rv64}.

\begin{figure}[ht]
  \centering
  \begin{lstlisting}[frame=single,breaklines=true,breakindent=0pt,basicstyle=\scriptsize\ttfamily,xleftmargin=0em]
Instruction encoding (bit fields):
| 31-20        | 19-15 | 14-12 | 11-7 | 6-0     |
| offset[11:0] | rs1   | 000   | rd   | 1100111 |

Format: jalr rd, rs1, offset
Description: Jump to address and place return address in rd.
Implementation: t = pc+4; pc = (x[rs1] + sext(offset)) & ~1; x[rd] = t
  \end{lstlisting}
  \caption{The requirement for the \texttt{JALR} instruction.}
  \label{fig:jalr-req}
\end{figure}

The user wants to implement \texttt{JALR} to the CPU emulator. 
\texttt{JALR} is the RISC-V indirect jump-and-link instruction:
it jumps to the address computed as \texttt{(x[rs1] + sext(offset)) \& \textasciitilde{}1},
storing the return address \texttt{pc+4} in \texttt{rd}.
Here \texttt{x[rs1]} denotes the value held in integer register \texttt{rs1},
and \texttt{sext(offset)} sign-extends the 12-bit immediate field to the full
register width.
Executing one \texttt{jalr rd, rs1, offset} touches two pieces of
architectural state and leaves the rest alone. The new
program counter is the sum of the sign-extended 12-bit
immediate and the value read from \texttt{rs1}, and with the least-significant bit cleared so that the
next fetch lands on a 2-byte boundary. The destination
register \texttt{rd} receives \texttt{pc + 4} as the return address. The other
thirty-one general-purpose registers, the byte-addressable
memory, the CSR file, the privilege mode, the halt flag, and
the host-event trace are all preserved.

In the Rocq instantiation, \approach{} processes the requirement in three
stages:

\begin{itemize}
    \item \textbf{Stage 1: Requirement analysis.} A
    requirement-analysis step turns the natural-language
    requirement into a \emph{coding plan} that lists the type
    definitions, functional definitions, and logic specifications
    the verified implementation must satisfy.

    \item \textbf{Stage 2: Synthesis and verification.} The
    \emph{coding agent} generates functional definitions for the pure
    core, and the \emph{proving agent} generates the corresponding
    specifications and proof scripts. With Rocq as the backend, these
    codes are Rocq definitions and proofs; rejected implementations are
    returned with compiler feedback for repair.

    \item \textbf{Stage 3: Composition and extraction.} \approach{} links
    the verified fragments, proves global properties of that pure
    core, and extracts it to the target language (C++ in our study).
    Host-language code implements effects such as I/O and runtime
    integration.
\end{itemize}

\subsection{Stage 1: Requirement analysis}

For the requirement in~\Cref{fig:jalr-req}, which paraphrases the
RISC-V instruction set manual, the requirement-analysis step
produces a coding plan with three parts.

The \emph{type definitions} fix how the concepts mentioned in the
requirement are represented as Rocq types and what target-language
types they map to after extraction. The architectural integer
width is pinned to \texttt{XLEN\_nat = 32}, so a machine word
is a 32-bit integer represented as \texttt{Z} in Rocq and
extracted to an \texttt{int64\_t} that always holds a value in
$[0, 2^{32})$. A register index and an immediate are also
\texttt{Z}. The register file is a \texttt{list word} of length
32 with \texttt{x0} pinned to zero by the convention that
\texttt{write\_reg} is a no-op at index 0. The whole machine is
a record \texttt{state} bundling the register file, the program
counter, the byte-addressable memory (an association list keyed
by addresses), the CSR file, a halt flag, the current privilege
mode, and a host-event trace. The plan keeps
them in the state so that the same record carries all
instructions.

The \emph{functional definitions} list the operations the
code must perform, together with their signatures. For
\texttt{JALR}, the plan asks for a single handler
\texttt{exec\_JALR : state $\to$ reg\_idx $\to$ reg\_idx $\to$
word $\to$ state} that takes the current machine state and the
three operands \texttt{rd}, \texttt{rs1}, \texttt{imm} and
returns the post-state.

The \emph{logic specifications} list the safety and correctness
properties the implementation must satisfy. The agent decomposes
the requirement into individually verifiable propositions whose
conjunction implies that the implementation matches the
requirement. For \texttt{JALR}, the plan produces three
function-level specifications: \texttt{rd} receives
\texttt{low\_XLEN(pc + 4)} (when \texttt{rd} is not \texttt{x0});
the new program counter equals
\texttt{jalr\_target(read\_reg(rs1) + sext\_from 12 imm)}, where
\texttt{jalr\_target} folds its argument mod $2^{32}$ and clears
bit~0; and bit~0 of the new program counter is always zero.

The coding plan also lists \emph{global specifications}:
properties the implementation must satisfy that no single
function definition expresses. For \texttt{JALR}, these are the
frame conditions on the parts of the state the instruction must
not perturb (memory, CSRs, mode, halt, trace, and every register
other than \texttt{rd}) together with the structural invariants
of the ISA as a whole: the register file stays length~32;
\texttt{x0} always reads as zero. Without these, an instruction that
writes the correct \texttt{rd} and pc but quietly corrupts an
unrelated register or shrinks the register file would still
violate correctness.

Figure~\ref{fig:jalr-coding-plan} shows the set of logic
specifications the coding plan emits for \texttt{JALR}.

\begin{figure}[ht]
\centering
\begin{lstlisting}[frame=single,breaklines=true,breakindent=0pt,
    basicstyle=\scriptsize\ttfamily,xleftmargin=0em]
Function-level specifications:
  jalr_link 
    rd receives low_bits XLEN (pc + 4) as the return address.
  jalr_pc  
    The new pc equals (x[rs1] + sext(offset)) with bit 0 cleared.
  jalr_lsb_zero
    Bit 0 of the new pc is always zero after exec_JALR.

Global specification:
  jalr_frame  [frame]
    Memory, CSRs, halt flag, privilege mode, and trace are unchanged. Every register other than rd is unchanged. The register file length is preserved.
\end{lstlisting}
\caption{The coding plan's logic specifications for \texttt{JALR}:
three function-level properties and one global invariant over the
composed \texttt{step} function.}
\label{fig:jalr-coding-plan}
\end{figure}


\subsection{Stage 2: Synthesis and verification}

The \texttt{JALR} handler reads from one register, writes to
another, and updates the program counter. To keep every handler
a pure, total Rocq term, \approach{} generates a set of common
helper functions that the coding agent uses to build handlers:
\texttt{read\_reg} and \texttt{write\_reg}
for the register file (the latter being a no-op at index~0),
\texttt{st\_regs\_pc} for the joint update of registers and pc,
\texttt{sext\_from\,$k$} for sign-extension from a $k$-bit
field, \texttt{addr\_mod} for the mod-$2^{32}$ address fold,
\texttt{low\_XLEN} for the XLEN-bit value fold, and
\texttt{jalr\_target} for the JALR-specific bit-0-clearing
operation. Each helper function is a pure function on \texttt{state}
or \texttt{Z}, so the resulting handler is a CIC term that
ordinary tactics can simplify and rewrite.

Working from the coding plan, the
coding agent emits the type definitions and the body of
\texttt{exec\_JALR}; the proving agent emits the formalized
version of each logic specification and a tactic script that
proves \texttt{exec\_JALR} satisfies it. Both outputs go to the
Rocq compiler. If the proof closes the goal, the generated code is
accepted and proceeds to extraction. Otherwise the compiler emits
an error, which \approach{} returns to the responsible agent together
with the local proof state and the source fragment the error
refers to. The agent then produces a revised proof, or a revised
implementation, and the
loop repeats. 
\Cref{fig:jalr} shows implementation and the logic specifications produced by the agent.

\begin{figure}[ht]
\begin{lstlisting}[language=Coq,basicstyle=\fontsize{8.2pt}{9.6pt}\selectfont\ttfamily]
(* Implementation *)
Definition exec_JALR (s : state) (rd rs1 : reg_idx) 
  (off : word) : state :=
    let link := low_XLEN (pc s + 4) in
    let summed := 
      read_reg (regs s) rs1 + sext_from 12 off in
    st_regs_pc s (write_reg (regs s) rd link) 
                 (jalr_target summed).

(* bit 0 of the new pc is always 0. *)
Lemma exec_JALR_lsb_zero :
  forall s rd rs1 off,
    Z.modulo (pc (exec_JALR s rd rs1 off)) 2 = 0.
Proof.
  intros s rd rs1 off.
  unfold exec_JALR, st_regs_pc.
  simpl pc.
  apply jalr_target_even.
Qed.
(* Global: regs always length 32. *)
Lemma global_reg_length : forall s i,
  length (regs s) = 32%nat ->
  length (regs (step s i)) = 32%nat.
Proof. ... Qed.
(* Global: x0 always reads as zero. *)
Lemma global_x0_zero : forall s i,
  read_reg (regs s) 0 = 0 ->
  read_reg (regs (step s i)) 0 = 0.
Proof. ... Qed.
...
\end{lstlisting}
\caption{The \texttt{exec\_JALR} implementation and the logic specifications produced by the agent.}
\label{fig:jalr}
\label{fig:jalr-specs}
\end{figure}

\subsection{Stage 3: Composition and extraction}
The verified \texttt{JALR} handler is one of handlers covering the given task requirements.
\approach{} links them into a single Rocq function \texttt{step} that
takes a state and a decoded instruction and dispatches to the
corresponding handler. 

\approach{} then proves composition theorems stating that
\texttt{step} preserves the global properties whenever
each constituent handler does: the register file stays length
32, \texttt{x0} stays zero, the program counter stays in $[0, 2^{32})$, non-CSR
instructions leave the CSR file alone, and non-memory
instructions leave memory alone. The theorems of composition are shown in~\Cref{fig:jalr-step-function}.

\begin{figure}[ht]
\begin{lstlisting}[language=Coq,basicstyle=\fontsize{8.2pt}{9.6pt}\selectfont\ttfamily,xleftmargin=0em]
Definition step (s : state) (i : instr) : state :=
  if halt s then s
  else
    match i with
    | LUI    rd imm     => 
        pc_advance4 (exec_LUI    s rd imm)
    | AUIPC  rd imm     =>
        pc_advance4 (exec_AUIPC  s rd imm)
    | ADDI   rd rs1 imm => 
        pc_advance4 (exec_ADDI   s rd rs1 imm)
    ...

(* The register file has exactly 32 entries. *)
Lemma global_reg_length :
  forall s i, length (regs s) = 32%nat ->
              length (regs (step s i)) = 32%nat.
Proof. ... Qed.
(* Non-CSR instructions leave the CSR unchanged. *)
Lemma global_non_csr_csrs_frame :
  forall s i, is_csr i = false -> halt s = false ->
              csrs (step s i) = csrs s.
Proof. ... Qed.
...
\end{lstlisting}
\caption{The \texttt{step} function and global specifications theorems for \texttt{step}.}
\label{fig:jalr-step-function}
\end{figure}

After all pure functions are verified, they are extracted to the
project's target language. The example uses Crane~\cite{bloomberg_crane_maintainers}
to emit C++, but the same workflow accepts other
extraction tools~\cite{10.1145/3656379,formal_land_coq_of_ocaml,formal_land_rocq_of_rust}
when the target is OCaml or Rust. 
In \approach{}, we assume the correctness of the extraction tool,
i.e., that it faithfully maps Rocq terms to the target language
without changing behaviors. Under this assumption, extraction preserves the
correctness properties proved inside Rocq, so the verified
functions can be deployed in projects written in different
languages without redoing the proof. 
For this interpreter, dispatch, instruction arithmetic, and modeled
architectural state updates go through the verified functions. The
Rocq-side guarantees therefore apply to those extracted definitions; the
host integration code is tested as part of the complete executable in
Section~\ref{sec:evaluation}.

\section{Approach}\label{sec:approach}

The three workflow stages in Section~\ref{sec:overview} are realized by a
requirement analyzer that produces coding plans, a coding agent and a proving
agent that generate and repair proof-assistant codes, and an extraction
step that translates accepted functional definitions into the target language.
The workflow is defined for an ITP backend; this section describes workflow used in \approach{}.

\subsection{Requirement analyzer}
\label{sec:coding-plan}
The requirement analyzer is an LLM agent that takes the natural-language
requirement as input and produces the coding plan.
It identifies the data representations required by the pure core, separates
pure computations from host-side effects, and decomposes each relevant
semantic clause into an independently verifiable property tied to a specific
function, including both what the function computes and what state it leaves
unchanged.

Because the coding plan must be unambiguous for synthesis to proceed, the analyzer treats any underspecified choice as a blocker rather than guessing a default.
When the requirement leaves a representational decision open---the target language, the integer width to use for a domain value, the signed or unsigned interpretation of a comparison---the analyzer issues a clarification question to the user and waits for an answer before continuing.
This dialogue repeats until target language, type choices, and logical properties in the coding plan are fully determined, at which point the plan is handed off to the coding and proving agents.

\noindent\textbf{Coding plan.}
A coding plan is a natural language document produced by the requirement analyzer from the requirement text.
It answers four questions: what data types represent the concepts in the
ITP language, what pure functions must be implemented, which operations
remain in the effectful host layer, and what logical properties the pure
implementation must satisfy.
The coding and proving agents both work from the coding plan rather than from
the raw requirement, so they share the same analysis of what the requirement
demands.

\noindent\textbf{Type definitions.}
The coding plan describes how each type
is represented in the selected ITP and what target-language type it
corresponds to after extraction.
In our Rocq backend, unbounded integers map to Rocq's \texttt{Z}, which
integrates with Rocq's integer arithmetic library; bounded values such as
array indices or byte offsets use \texttt{Z} together with a range predicate.
Finite sequences of homogeneous elements map to \texttt{list}, with a length invariant added when the size is fixed by the requirement.
Product types become Rocq \texttt{Record} definitions in this backend, with
fields corresponding to the named components in the requirement.
The extraction target for each Rocq type is determined by the project's target language and the requirement.
\texttt{Z} extracts to \texttt{int64\_t} in a C or C++ project, to \texttt{i64} in a Rust project, and to \texttt{int} or \texttt{int64} in an OCaml project; a length-constrained \texttt{list Z} becomes the corresponding fixed-size array type in each language.
Recording these correspondences explicitly in the coding plan ensures that both agents and the extraction backend agree on the same types, and any mismatch surfaces as a type error before synthesis produces incorrect code.

\noindent\textbf{Functional definitions.}
The coding plan distinguishes deterministic computations that belong in the
ITP-defined from operations that must remain in the
target-language host code.
Any deterministic computation whose correctness is asserted by the
requirement---core logic, modeled state transformations, or arithmetic over
domain values---is assigned to a pure function so that the proving agent can
reason about it.
Operations such as concrete I/O, allocation, and command-line handling are
recorded as host obligations and implemented in the target language; they
are not claimed to be verified by the pure-core proofs.

When a requirement involves multiple distinct computations, the coding plan
assigns each one to a separate pure function.
Keeping functions small and focused means the proving agent can verify each in isolation, and the correctness of the whole follows by composition.
The composed functions also require theorems stating the correctness of their
composition.

\noindent\textbf{Logical properties.}
The coding plan lists a correctness property for each function.
Each property has two parts: a computes clause that states what the function's output must equal, and global properties that list the state fields the function must not change.

\subsection{Coding Agent}
\label{sec:coding-agent}

The coding agent receives the coding plan produced by the requirement analyzer
as its context and produces type declarations and pure function
implementations for the selected ITP backend. 

\noindent\textbf{Coding plan as a scaffold.}
Working from the coding plan, the coding agent synthesizes two kinds of
artifacts: type declarations using \texttt{Inductive} or \texttt{Record} for
each concept in the type definitions section, and function bodies using
\texttt{Definition} or \texttt{Fixpoint} for each entry in the functional
definitions section.
The coding plan constrains this synthesis in two ways.
The type definitions section fixes the representation for every concept: the agent cannot invent an alternative type for something already assigned one.
The functional definitions section fixes each function's type signature and its corresponding semantic clause, so the agent focuses on the body rather than choosing a type-compatible but semantically different contract.

\subsection{Proving Agent}
\label{sec:proving-agent}

The proving agent receives the logical properties section of the coding plan
and the accepted type definitions and function implementations from the
coding agent.
For each property, it produces a proposition in the selected ITP and a proof
that checks the generated pure implementation. 

\noindent\textbf{Property formalization.}
Each natural language property in the coding plan is translated into a theorem using the type from the type definitions and the function from the coding agent's output.
By expressing these properties as propositions and requiring the ITP's
kernel to accept a proof of each one, the ITP backend instantiation establishes the
specified properties of the synthesized pure functions through checked
proofs rather than testing alone.

\noindent\textbf{Tactic generation.}
For each theorem, the proving agent generates a proof script---a sequence of proof steps such as \texttt{intros}, \texttt{unfold}, \texttt{rewrite}, and \texttt{reflexivity}---that attempts to reduce the proof goal to \texttt{True}.
When the ITP accepts the script, every subgoal has been discharged and the theorem is established as a machine-checked fact.

Proof-guaranteed correctness is the core mechanism by which \approach{} eliminates LLM hallucinations.
LLMs may generate code that compiles and passes tests yet violates a semantic property the requirement demands.
\approach{} counters this by asking the LLM not only to synthesize the function but also to state the properties the function must satisfy and to produce a proof script that establishes each property against the synthesized definition.
These two kinds of codes---the implementation and its proof---are submitted together to the ITP backend.
If the implementation contains an error, the proof cannot go through: the compiler will find a subgoal it cannot close, no matter how convincingly the proof script is written.
There is no way for the LLM to hallucinate a proof into acceptance; the compiler either finds a closed derivation or rejects the submission.
Conversely, if the compiler accepts the proof, the property is established as a mathematical truth, and the synthesized function is guaranteed to satisfy it for every input it can ever receive.
The hallucination is therefore not suppressed by prompting or filtered by post-hoc analysis---it is structurally impossible for a hallucinated implementation to pass verification while violating its specification.

\subsection{Repair Loop}
\label{sec:repair}

In the Rocq instantiation, when the compiler rejects a generated definition
or proof, \approach{} returns the error and relevant context to the
responsible agent for repair.

\noindent\textbf{Code errors.}
Errors in the coding agent's output fall into two categories.
The first is type errors: a type mismatch, an undefined identifier, or an incorrect number of arguments.
The compiler message names the error definition and its expected type, giving the agent a precise location to rewrite; the repair requires changing a single term rather than rewriting the whole definition.
The second is semantic errors in the function body: the definition computes the wrong result, which surfaces when the proving agent cannot discharge a specification against it.
In this case, the coding agent receives the coding plan, the current function body, and the error feedback, and uses them together to diagnose and repair the definition.

\noindent\textbf{Proof errors.}
A proof error occurs when the compiler reports a tactic failure or finds a subgoal still open after the script's final step.
The proving agent receives the error together with the Rocq proof state at the failure point and attempts to repair the tactic script.
If the repaired script closes the proof, the theorem is established.
If the repair fails, the agent tries again, revising the script based on each new error report.
Once $k$ repair attempts all fail, the agent stops patching the script and reflects: it re-examines the theorem statement, the function body, and the type definitions against the original requirement to verify that each code faithfully represents what the requirement demands.
If any code is found to deviate from the requirement, the agent regenerates that code before attempting a new proof; otherwise it proceeds to synthesizing a new proof script from scratch.
If the new script also fails within the threshold, the proof is considered beyond the proving agent's capability.
\approach{} surfaces the failing goal, the last attempted proof script, and the current function body to the user with a prompt requesting manual intervention to complete or guide the proof.
Even in this case the user's effort is substantially less than in conventional verified development: the functional code and all other proof scripts have already been synthesized and accepted, so the user need only address the specific failing goal rather than writing the complete implementation and all correctness proofs.

\subsection{Extraction}
\label{sec:extraction}
Extraction targets only the computational content of the verified core. In
the Rocq backend, these are the \texttt{Inductive}, \texttt{Record},
\texttt{Definition}, and \texttt{Fixpoint} declarations that constitute the
implementation; theorems and proof terms are erased during extraction.
\approach{} assumes that the extraction backend faithfully preserves the
operational semantics of the accepted source definitions.

The host program supplies concrete implementations for the effects excluded
from the pure core, such as I/O routines and other runtime integration. These
host-side components and the extracted definitions form the complete,
runnable project. 

\section{Evaluation}\label{sec:evaluation}

We evaluate \approach{} on three research questions:

\begin{description}
  \item[\textbf{RQ1}] Can an ITP-backed workflow construct a complete,
    non-trivial executable project with a verified functional core?
  \item[\textbf{RQ2}] How does an ITP backend compare with an
    automated-verification backend on the same project-generation task?
  \item[\textbf{RQ3}] Does the integrated project behave robustly under
    generated tests and fuzzing after extraction?
\end{description}

\subsection{Setup}

\noindent\textbf{Implementation.}
We build \approach{} on top of Cursor Agent~\cite{cursor_agents}, which drives
all three roles in the pipeline: the Requirement Analyzer, the Coding
Agent, and the Proving Agent.
For the verification backend we use two tools, Rocq~9.0.1~\cite{CoqRefMan900}
and Dafny~4.11.0~\cite{DBLP:conf/lpar/Leino10}, representing two different
ways to supply a verified core. Rocq represents the ITP paradigm (e.g.,
Lean~\cite{DBLP:conf/cade/Moura021}, Isabelle~\cite{DBLP:conf/tphol/WenzelPN08}),
where explicit proofs are checked by a trusted kernel; Dafny represents an
automated-verification paradigm (e.g.,
Verus~\cite{lattuada2023verus}), where specifications are
embedded in code and verification conditions are proved by an automated
solver. In each configuration, successful verification is followed by
generation of target-language code for integration.

\noindent\textbf{Configuration.}
RQ1 and RQ3 use Cursor Agent with Rocq~9.0.1 as the \approach{}
backend; RQ2 repeats the synthesis task with Dafny~4.11.0 as a comparative
backend.
All experiments were conducted on a server equipped with four 20-core Intel Ultra 265k, 48 GB RAM, and running Ubuntu 22.04.5 LTS. 

\subsection{RQ1: Can an ITP-backed workflow construct a complete, non-trivial executable project with a verified functional core?}
\textbf{Procedure.}
To balance project scale against experimental cost, we ask \approach{} to
synthesize an interpreter for the unprivileged RV32I base: 47 instructions
covering integer arithmetic and logic, shifts, loads and stores,
conditional branches, the jumps, memory accesses, and the environment calls.
Each instruction enters the pipeline as one natural-language requirement
quoted from the RISC-V unprivileged ISA manual~\cite{riscv_unpriv_rv64}.
\approach{} emits, per instruction, a Rocq handler with its function-level and
global specifications and the proof scripts that prove them.
We extract the result to C++, recording the size of the synthesized
Rocq sources and the extracted C++ sources in lines of code.
The run begins with the requirement texts and target configuration. After
launch, no human reviews or edits the Rocq definitions,
specifications, proof scripts, repair attempts, extracted output, or C++
host-layer glue code; project construction is performed end-to-end by the
agent.
All synthesis runs use Claude Opus 4.7 High as the underlying model,
with a 30-minute timeout.

\begin{table}[htbp]
  \centering
  \caption{Rocq and C++ source line counts}
  \label{tab:coq-loc}
  \begin{tabular}{lr}
    \toprule
    Category & Lines \\
    \midrule
    \texttt{Function Definitions (Rocq)}            & 821 \\
    \texttt{Types (Rocq)}                & 116 \\
    \texttt{Properties (Rocq)}           & 922 \\
    \midrule
    \textbf{Total (Rocq)}                  & \textbf{1,859} \\
    \midrule
    \texttt{Extracted code (C++)}            & 2,848 \\
    \texttt{Host project (C++)}            & 88 \\
    \midrule
    \textbf{Total (C++)}            & 2,936 \\
    \bottomrule
  \end{tabular}
\end{table}

\textbf{Results.}
The full synthesis consumed approximately 21 million tokens in total.
Table~\ref{tab:coq-loc} shows the size of the synthesized Rocq sources and the extracted C++ sources.
\approach{} automatically synthesized a coding plan for each of the 47
instructions and, from those plans, generated 1,859 lines of Rocq: 116
lines of type definitions, 821 lines of function definitions, and 922
lines of specifications and proof scripts.
The proof code is kept compact because the LLM synthesized reusable Ltac
tactics shared across multiple theorems.
\approach{} then extracted 2,848 lines of C++ from the verified Rocq sources;
the host project contributes only 88 lines of glue code for I/O and runtime
integration.
Thus the interpreter and its verified functional core are produced without
human intervention during development.

\begin{rqbox}
  \textbf{Answer to RQ1:}
  In this case study, an LLM agent paired with an ITP backend fully
  automatically constructs a complete non-trivial project within 30
  minutes, with its functional core verified in Rocq and extracted into a
  runnable C++ project.
\end{rqbox}

\subsection{RQ2: How does an ITP backend compare with an automated-verification backend on the same project-generation task?}
\textbf{Procedure.}
We replace \approach{}'s Rocq backend with Dafny~\cite{DBLP:conf/lpar/Leino10}, a
programming language that supports verification through pre- and
postconditions written inline alongside C-style code.
Unlike Rocq, which requires explicit tactic proofs, Dafny discharges
proof obligations automatically via its SMT backend; the verified code
can then be exported to other languages.
We also adapt the coding plan generation accordingly: the requirement
analyzer is updated to emit Dafny-compatible type declarations,
function signatures, and pre/postcondition templates in place of Rocq
type and logical property descriptions.
We run the same synthesis task as RQ1.

\textbf{Results.}
The LLM agent failed to complete the synthesis task within the
30-minute timeout.
It generated 2,070 lines of Dafny code during that time but did not
produce a complete, verified set of pre- and postconditions for all
functions.
The observed failure mode is a feedback gap in the Dafny verification loop:
when the solver times out on verification conditions in this run, it returns
no useful information about why the obligation was not discharged.
Without a useful local error message, the agent cycles through alternative
verification conditions, each triggering another solver timeout, until the
experiment reaches its time budget. In contrast, a Rocq tactic failure in
our run produces a concrete proof state showing which subgoal remains open,
giving the proving agent a precise target for its next repair attempt.

\begin{rqbox}
  \textbf{Answer to RQ2:}
  Under the same task and 30-minute budget, the Rocq backend completes the
  interpreter while the Dafny backend does not complete verification. 
  Explicit proof construction makes Rocq better suited to LLM-agent-driven
  development than approaches based on pre/postconditions and SMT solving.
\end{rqbox}

\subsection{RQ3: Does the integrated project behave robustly under generated tests and fuzzing after extraction?}
\textbf{Procedure.}
To dynamically validate the integrated C++ interpreter, including behavior
across its unverified host boundary, we apply two complementary methods.
First, we use the LLM to generate test cases for each instruction and
run them against the extracted code.
Second, we feed the LLM-generated test cases as initial seeds into
AFL++~\cite{AFLplusplus-Woot20} and test for 12 hours.

\textbf{Results.}
The LLM generated 265 test cases covering all 47 instructions; all 265
passed against the extracted C++ interpreter.
The AFL++ campaign ran for 12 hours and executed 98.2 million inputs, completing 211 full fuzzing cycles.
It found zero crashes and zero hangs.

\begin{rqbox}
  \textbf{Answer to RQ3:}
  Although the small host C++ layer handling side effects is unverified,
  the combined evidence from 265 passing instruction-level tests and 12
  hours of AFL++ fuzzing with no crashes or hangs supports the feasibility
  of integrating the extracted verified core into a runnable project. 
\end{rqbox}

\section{Related Work}\label{sec:relation}

\textbf{LLM-based code generation.}
LLMs including GPT~\cite{achiam2023gpt}, Claude~\cite{anthropic2026sonnet46},
and DeepSeek-Coder~\cite{guo2024deepseekcoderlargelanguagemodel,deepseekai2025deepseekv32}
perform strongly on code-generation and software-engineering benchmarks such
as HumanEval~\cite{chen2021evaluatinglargelanguagemodels} and
SWE-bench~\cite{jimenez2024swebench}. Coding agents such as
Cursor~\cite{research2026composer2technicalreport}, Claude
Code~\cite{claudecode2025}, and OpenCode~\cite{opencode} extend these models
with editing, execution, and testing tools. Such tools make natural-language
project development practical, but generated code can still pass available
tests while deviating from intended
semantics~\cite{li2026fmagent,DBLP:journals/tois/HuangYMZFWCPFQL25}. Our study
addresses this reliability concern by evaluating a workflow in which the
project's core is subject to checked proofs rather than testing alone.

\textbf{Verification with ITPs.}
Rocq~\cite{CoqRefMan900,bertot2013interactive},
Lean~4~\cite{DBLP:conf/cade/Moura021}, and
Isabelle/HOL~\cite{DBLP:conf/tphol/WenzelPN08} have served as trust tools for
safety-critical systems such as seL4~\cite{DBLP:conf/sosp/KleinEHACDEEKNSTW09},
CertiKOS~\cite{DBLP:conf/osdi/GuSCWKSC16}, and
CompCert~\cite{Leroy-BKSPF-2016}, though each required person-years of manual proof
effort. Several lines of work reduce that cost:
CoqHammer~\cite{DBLP:journals/jar/CzajkaK18},
Sledgehammer~\cite{DBLP:conf/cade/BohmeN10}, and
Lean-auto~\cite{qian2025leanautointerfacelean4} discharge subgoals to SMT solvers,
while CoqGym~\cite{DBLP:conf/icml/YangD19},
Proverbot9001~\cite{DBLP:conf/pldi/Sanchez-SternAS20}, and
Graph2Tac~\cite{DBLP:conf/icml/BlaauwbroekORMP24} train neural models to predict
the next tactic.
The goal of \approach{} is to reduce the development
cost of large safety-critical systems by leveraging LLM agents' code-generation ability
to produce the Rocq and other programming languages codes that the project expects.

\textbf{LLM agents for automated theorem proving.}
Now many LLM agents for automated theorem proving have been proposed.
The agent sees the proof state, proposes a tactic, and iterates on compiler feedback.
LeanDojo~\cite{DBLP:conf/nips/YangSGCSYGPA23} uses retrieval-augmented generation
over mathlib to predict the next tactic at scale.
LEGO-Prover~\cite{DBLP:conf/iclr/WangXZLCHXSX0LL24} accumulates a reusable lemma
library across proof tasks.
On the Rocq,
PALM~\cite{DBLP:conf/kbse/LuD024} combine LLMs with error-driven
repair loops.
Cobblestone~\cite{DBLP:journals/corr/abs-2410-19940} uses divide-and-conquer strategies to decompose a proof into smaller subgoals. 
Rango~\cite{DBLP:conf/icse/ThompsonSCFSB0L25} retrieves the lemmas and proof scripts
most relevant to the current goal.

All of these assume the specification is given and target only the proof.
\approach{} starts one step earlier: the user provides a natural-language
requirement, and the coding agent and proving agent jointly produce the
implementation, the formal specification, and the proof, which verified extraction
then carries into the host-language integration.
\section{Conclusion}

This paper investigates whether interactive theorem provers provide a viable
backend for LLM-assisted verified programming at software-project scale. We
instantiate an ITP-based workflow with Rocq: the agent separates pure logic
from host-side effects, generates and proves the pure functional core, and
extracts that core into C++ for integration with a runnable project. 

The case study provides empirical evidence of feasibility. For a complete
interpreter covering 47 instructions of the unprivileged RISC-V RV32I
base, the Rocq-backed agent completes within 30 minutes, generates 1,859
lines of verified Rocq, and extracts 2,848 lines of C++. The resulting
interpreter passes 265 LLM-generated tests and completes 12 hours of AFL++
fuzzing with zero crashes and zero hangs.

The matched Dafny run does not complete verification within the same
30-minute budget. Rocq's explicit proof states
provide local repair targets, whereas solver timeouts encountered in the
Dafny run do not provide similarly actionable feedback. This observation
suggests that feedback quality is an important design consideration when
verification backends are operated by LLM agents.

Taken together, the results establish that ITP-based verified programming is
a credible route for LLM-generated software projects, rather than only for
isolated theorem-proving tasks.

%
\IEEEpeerreviewmaketitle

\bibliographystyle{IEEEtran}
\bibliography{references}

@misc{guo2024deepseekcoderlargelanguagemodel,
      title={DeepSeek-Coder: When the Large Language Model Meets Programming -- The Rise of Code Intelligence}, 
      author={Daya Guo and Qihao Zhu and Dejian Yang and Zhenda Xie and Kai Dong and Wentao Zhang and Guanting Chen and Xiao Bi and Y. Wu and Y. K. Li and Fuli Luo and Yingfei Xiong and Wenfeng Liang},
      year={2024},
      eprint={2401.14196},
      archivePrefix={arXiv},
      primaryClass={cs.SE},
      url={https://arxiv.org/abs/2401.14196}, 
}

@misc{chen2021evaluatinglargelanguagemodels,
      title={Evaluating Large Language Models Trained on Code}, 
      author={Mark Chen and Jerry Tworek and Heewoo Jun and Qiming Yuan and Henrique Ponde de Oliveira Pinto and Jared Kaplan and Harri Edwards and Yuri Burda and Nicholas Joseph and Greg Brockman and Alex Ray and Raul Puri and Gretchen Krueger and Michael Petrov and Heidy Khlaaf and Girish Sastry and Pamela Mishkin and Brooke Chan and Scott Gray and Nick Ryder and Mikhail Pavlov and Alethea Power and Lukasz Kaiser and Mohammad Bavarian and Clemens Winter and Philippe Tillet and Felipe Petroski Such and Dave Cummings and Matthias Plappert and Fotios Chantzis and Elizabeth Barnes and Ariel Herbert-Voss and William Hebgen Guss and Alex Nichol and Alex Paino and Nikolas Tezak and Jie Tang and Igor Babuschkin and Suchir Balaji and Shantanu Jain and William Saunders and Christopher Hesse and Andrew N. Carr and Jan Leike and Josh Achiam and Vedant Misra and Evan Morikawa and Alec Radford and Matthew Knight and Miles Brundage and Mira Murati and Katie Mayer and Peter Welinder and Bob McGrew and Dario Amodei and Sam McCandlish and Ilya Sutskever and Wojciech Zaremba},
      year={2021},
      eprint={2107.03374},
      archivePrefix={arXiv},
      primaryClass={cs.LG},
      url={https://arxiv.org/abs/2107.03374}, 
}

@online{claude_code_overview,
  author       = {{Anthropic}},
  title        = {Claude Code Documentation: Overview},
  year         = {2026},
  url          = {https://code.claude.com/docs/en/overview},
  urldate      = {2026-04-27},
  organization = {Anthropic}
}

@online{cursor_agents,
  author       = {{Anysphere}},
  title        = {Cursor Agents},
  year         = {2026},
  organization = {Anysphere}
}

@online{opencode,
  author       = {{OpenCode Team}},
  title        = {OpenCode: Open source, terminal-based {AI} coding assistant},
  year         = {2025},
  url          = {https://opencode.ai},
  organization = {Anomaly}
}

@misc{li2026fmagent,
      title={The FM Agent}, 
      author={Annan Li and Chufan Wu and Zengle Ge and Yee Hin Chong and Zhinan Hou and Lizhe Cao and Cheng Ju and Jianmin Wu and Huaiming Li and Haobo Zhang and Shenghao Feng and Mo Zhao and Fengzhi Qiu and Rui Yang and Mengmeng Zhang and Wenyi Zhu and Yingying Sun and Quan Sun and Shunhao Yan and Danyu Liu and Dawei Yin and Dou Shen},
      year={2026},
      eprint={2510.26144},
      archivePrefix={arXiv},
      primaryClass={cs.AI},
      url={https://arxiv.org/abs/2510.26144}, 
}

@article{DBLP:journals/tois/HuangYMZFWCPFQL25,
  author       = {Lei Huang and
                  Weijiang Yu and
                  Weitao Ma and
                  Weihong Zhong and
                  Zhangyin Feng and
                  Haotian Wang and
                  Qianglong Chen and
                  Weihua Peng and
                  Xiaocheng Feng and
                  Bing Qin and
                  Ting Liu},
  title        = {A Survey on Hallucination in Large Language Models: Principles, Taxonomy,
                  Challenges, and Open Questions},
  journal      = {{ACM} Trans. Inf. Syst.},
  volume       = {43},
  number       = {2},
  pages        = {42:1--42:55},
  year         = {2025},
  url          = {https://doi.org/10.1145/3703155},
  doi          = {10.1145/3703155},
  bibsource    = {dblp computer science bibliography, https://dblp.org}
}

@inproceedings{Leroy-BKSPF-2016,
  author = {Xavier Leroy and Sandrine Blazy and Daniel K\"astner
            and Bernhard Schommer and Markus Pister and Christian Ferdinand},
  title = {CompCert -- A Formally Verified Optimizing Compiler},
  booktitle = {ERTS 2016: Embedded Real Time Software and Systems},
  publisher = {SEE},
  year = 2016,
  url = {http://xavierleroy.org/publi/erts2016_compcert.pdf},
  hal = {https://hal.inria.fr/hal-01238879},
  xtopic = {compcert}
}

@inproceedings{DBLP:conf/sosp/KleinEHACDEEKNSTW09,
  author       = {Gerwin Klein and
                  Kevin Elphinstone and
                  Gernot Heiser and
                  June Andronick and
                  David A. Cock and
                  Philip Derrin and
                  Dhammika Elkaduwe and
                  Kai Engelhardt and
                  Rafal Kolanski and
                  Michael Norrish and
                  Thomas Sewell and
                  Harvey Tuch and
                  Simon Winwood},
  editor       = {Jeanna Neefe Matthews and
                  Thomas E. Anderson},
  title        = {se{L}4: formal verification of an OS kernel.},
  booktitle    = {Proceedings of the 22nd {ACM} Symposium on Operating Systems Principles
                  2009, {SOSP} 2009, Big Sky, Montana, USA, October 11-14, 2009},
  pages        = {207--220},
  publisher    = {{ACM}},
  year         = {2009},
  url          = {https://doi.org/10.1145/1629575.1629596},
  doi          = {10.1145/1629575.1629596},
  bibsource    = {dblp computer science bibliography, https://dblp.org}
}

@inproceedings{DBLP:conf/osdi/GuSCWKSC16,
  author       = {Ronghui Gu and
                  Zhong Shao and
                  Hao Chen and
                  Xiongnan (Newman) Wu and
                  Jieung Kim and
                  Vilhelm Sj{\"{o}}berg and
                  David Costanzo},
  editor       = {Kimberly Keeton and
                  Timothy Roscoe},
  title        = {Certi{KOS}: An Extensible Architecture for Building Certified Concurrent OS Kernels},
  booktitle    = {12th {USENIX} Symposium on Operating Systems Design and Implementation,
                  {OSDI} 2016, Savannah, GA, USA, November 2-4, 2016},
  pages        = {653--669},
  publisher    = {{USENIX} Association},
  year         = {2016},
  url          = {https://www.usenix.org/conference/osdi16/technical-sessions/presentation/gu},
  bibsource    = {dblp computer science bibliography, https://dblp.org}
}

@misc{CoqRefMan900,
  title = {The {Coq} Reference Manual -- Release 9.0},
  author = {{The Coq Development Team}},
  year = {2025},
  url = {https://rocq-prover.org/doc/v9.0/refman/index.html},
}

@inproceedings{DBLP:conf/tphol/WenzelPN08,
  author       = {Makarius Wenzel and
                  Lawrence C. Paulson and
                  Tobias Nipkow},
  editor       = {Otmane A{\"{\i}}t Mohamed and
                  C{\'{e}}sar A. Mu{\~{n}}oz and
                  Sofi{\`{e}}ne Tahar},
  title        = {The Isabelle Framework},
  booktitle    = {Theorem Proving in Higher Order Logics, 21st International Conference,
                  TPHOLs 2008, Montreal, Canada, August 18-21, 2008. Proceedings},
  series       = {Lecture Notes in Computer Science},
  volume       = {5170},
  pages        = {33--38},
  publisher    = {Springer},
  year         = {2008},
  url          = {https://doi.org/10.1007/978-3-540-71067-7\_7},
  doi          = {10.1007/978-3-540-71067-7\_7},
  bibsource    = {dblp computer science bibliography, https://dblp.org}
}

@inproceedings{DBLP:conf/cade/Moura021,
  author       = {Leonardo de Moura and
                  Sebastian Ullrich},
  editor       = {Andr{\'{e}} Platzer and
                  Geoff Sutcliffe},
  title        = {The Lean 4 Theorem Prover and Programming Language},
  booktitle    = {Automated Deduction - {CADE} 28 - 28th International Conference on
                  Automated Deduction, Virtual Event, July 12-15, 2021, Proceedings},
  series       = {Lecture Notes in Computer Science},
  volume       = {12699},
  pages        = {625--635},
  publisher    = {Springer},
  year         = {2021},
  url          = {https://doi.org/10.1007/978-3-030-79876-5\_37},
  doi          = {10.1007/978-3-030-79876-5\_37},
  bibsource    = {dblp computer science bibliography, https://dblp.org}
}

@inproceedings{DBLP:conf/icse/ThompsonSCFSB0L25,
  author       = {Kyle Thompson and
                  Nuno Saavedra and
                  Pedro Carrott and
                  Kevin Fisher and
                  Alex Sanchez{-}Stern and
                  Yuriy Brun and
                  Jo{\~{a}}o F. Ferreira and
                  Sorin Lerner and
                  Emily First},
  title        = {Rango: Adaptive Retrieval-Augmented Proving for Automated Software
                  Verification},
  booktitle    = {47th {IEEE/ACM} International Conference on Software Engineering,
                  {ICSE} 2025, Ottawa, ON, Canada, April 26 - May 6, 2025},
  pages        = {347--359},
  publisher    = {{IEEE}},
  year         = {2025},
  url          = {https://doi.org/10.1109/ICSE55347.2025.00161},
  doi          = {10.1109/ICSE55347.2025.00161},
  bibsource    = {dblp computer science bibliography, https://dblp.org}
}

@inproceedings{DBLP:conf/kbse/LuD024,
  author       = {Minghai Lu and
                  Benjamin Delaware and
                  Tianyi Zhang},
  editor       = {Vladimir Filkov and
                  Baishakhi Ray and
                  Minghui Zhou},
  title        = {Proof Automation with Large Language Models},
  booktitle    = {Proceedings of the 39th {IEEE/ACM} International Conference on Automated
                  Software Engineering, {ASE} 2024, Sacramento, CA, USA, October 27
                  - November 1, 2024},
  pages        = {1509--1520},
  publisher    = {{ACM}},
  year         = {2024},
  url          = {https://doi.org/10.1145/3691620.3695521},
  doi          = {10.1145/3691620.3695521},
  bibsource    = {dblp computer science bibliography, https://dblp.org}
}

@article{DBLP:journals/corr/abs-2410-19940,
  author       = {Saketh Ram Kasibatla and
                  Arpan Agarwal and
                  Yuriy Brun and
                  Sorin Lerner and
                  Talia Ringer and
                  Emily First},
  title        = {Cobblestone: Iterative Automation for Formal Verification},
  journal      = {CoRR},
  volume       = {abs/2410.19940},
  year         = {2024},
  url          = {https://doi.org/10.48550/arXiv.2410.19940},
  doi          = {10.48550/ARXIV.2410.19940},
  eprinttype    = {arXiv},
  eprint       = {2410.19940},
  bibsource    = {dblp computer science bibliography, https://dblp.org}
}

@book{bertot2013interactive,
  title={Interactive theorem proving and program development: Coq’Art: the calculus of inductive constructions},
  author={Bertot, Yves and Cast{\'e}ran, Pierre},
  year={2013},
  publisher={Springer Science \& Business Media}
}

@article{DBLP:journals/jar/CzajkaK18,
  author       = {Lukasz Czajka and
                  Cezary Kaliszyk},
  title        = {Hammer for Coq: Automation for Dependent Type Theory},
  journal      = {J. Autom. Reason.},
  volume       = {61},
  number       = {1-4},
  pages        = {423--453},
  year         = {2018},
  url          = {https://doi.org/10.1007/s10817-018-9458-4},
  doi          = {10.1007/S10817-018-9458-4},
  bibsource    = {dblp computer science bibliography, https://dblp.org}
}

@misc{deepseekai2025deepseekv32,
      title={DeepSeek-V3.2: Pushing the Frontier of Open Large Language Models}, 
      author={DeepSeek-AI},
      year={2025},
}

@inproceedings{DBLP:conf/icml/YangD19,
  author       = {Kaiyu Yang and
                  Jia Deng},
  editor       = {Kamalika Chaudhuri and
                  Ruslan Salakhutdinov},
  title        = {Learning to Prove Theorems via Interacting with Proof Assistants},
  booktitle    = {Proceedings of the 36th International Conference on Machine Learning,
                  {ICML} 2019, 9-15 June 2019, Long Beach, California, {USA}},
  series       = {Proceedings of Machine Learning Research},
  volume       = {97},
  pages        = {6984--6994},
  publisher    = {{PMLR}},
  year         = {2019},
  url          = {http://proceedings.mlr.press/v97/yang19a.html},
  bibsource    = {dblp computer science bibliography, https://dblp.org}
}

@inproceedings{DBLP:conf/icml/BlaauwbroekORMP24,
  author       = {Lasse Blaauwbroek and
                  Mirek Ols{\'{a}}k and
                  Jason Rute and
                  Fidel Ivan Schaposnik Massolo and
                  Jelle Piepenbrock and
                  Vasily Pestun},
  title        = {Graph2Tac: Online Representation Learning of Formal Math Concepts},
  booktitle    = {Forty-first International Conference on Machine Learning, {ICML} 2024,
                  Vienna, Austria, July 21-27, 2024},
  publisher    = {OpenReview.net},
  year         = {2024},
  url          = {https://openreview.net/forum?id=A7CtiozznN},
  bibsource    = {dblp computer science bibliography, https://dblp.org}
}

@inproceedings{DBLP:conf/iclr/WangXZLCHXSX0LL24,
  author       = {Haiming Wang and
                  Huajian Xin and
                  Chuanyang Zheng and
                  Zhengying Liu and
                  Qingxing Cao and
                  Yinya Huang and
                  Jing Xiong and
                  Han Shi and
                  Enze Xie and
                  Jian Yin and
                  Zhenguo Li and
                  Xiaodan Liang},
  title        = {LEGO-Prover: Neural Theorem Proving with Growing Libraries},
  booktitle    = {The Twelfth International Conference on Learning Representations,
                  {ICLR} 2024, Vienna, Austria, May 7-11, 2024},
  publisher    = {OpenReview.net},
  year         = {2024},
  url          = {https://openreview.net/forum?id=3f5PALef5B},
  bibsource    = {dblp computer science bibliography, https://dblp.org}
}

@inproceedings{DBLP:conf/pldi/Sanchez-SternAS20,
  author       = {Alex Sanchez{-}Stern and
                  Yousef Alhessi and
                  Lawrence K. Saul and
                  Sorin Lerner},
  editor       = {Koushik Sen and
                  Mayur Naik},
  title        = {Generating correctness proofs with neural networks},
  booktitle    = {Proceedings of the 4th {ACM} {SIGPLAN} International Workshop on Machine
                  Learning and Programming Languages, MAPL@PLDI 2020, London, UK, June
                  15, 2020},
  pages        = {1--10},
  publisher    = {{ACM}},
  year         = {2020},
  url          = {https://doi.org/10.1145/3394450.3397466},
  doi          = {10.1145/3394450.3397466},
  bibsource    = {dblp computer science bibliography, https://dblp.org}
}

@misc{qian2025leanautointerfacelean4,
      title={Lean-auto: An Interface between Lean 4 and Automated Theorem Provers}, 
      author={Yicheng Qian and Joshua Clune and Clark Barrett and Jeremy Avigad},
      year={2025},
      eprint={2505.14929},
      archivePrefix={arXiv},
      primaryClass={cs.LO},
      url={https://arxiv.org/abs/2505.14929}, 
}

@inproceedings{DBLP:conf/cade/BohmeN10,
  author       = {Sascha B{\"{o}}hme and
                  Tobias Nipkow},
  editor       = {J{\"{u}}rgen Giesl and
                  Reiner H{\"{a}}hnle},
  title        = {Sledgehammer: Judgement Day},
  booktitle    = {Automated Reasoning, 5th International Joint Conference, {IJCAR} 2010,
                  Edinburgh, UK, July 16-19, 2010. Proceedings},
  series       = {Lecture Notes in Computer Science},
  volume       = {6173},
  pages        = {107--121},
  publisher    = {Springer},
  year         = {2010},
  url          = {https://doi.org/10.1007/978-3-642-14203-1\_9},
  doi          = {10.1007/978-3-642-14203-1\_9},
  bibsource    = {dblp computer science bibliography, https://dblp.org}
}

@misc{claudecode2025,
  author       = {Anthropic},
  title        = {Claude Code: An agentic coding tool},
  howpublished = {\url{https://code.claude.com/}},
  year         = {2025},
}

@misc{paraskevopoulou2026machinegeneratedmachinecheckedproofsverified,
      title={Machine-Generated, Machine-Checked Proofs for a Verified Compiler (Experience Report)}, 
      author={Zoe Paraskevopoulou},
      year={2026},
      eprint={2602.20082},
      archivePrefix={arXiv},
      primaryClass={cs.PL},
      url={https://arxiv.org/abs/2602.20082}, 
}

@misc{anthropic2026sonnet46,
  author       = {Anthropic},
  title        = {Introducing Claude Sonnet 4.6},
  howpublished = {\url{https://www.anthropic.com/news/claude-sonnet-4-6}},
  year         = {2026},
}

@inproceedings{anand2017certicoq,
  title={CertiCoq: A verified compiler for Coq},
  author={Anand, Abhishek and Appel, Andrew and Morrisett, Greg and Paraskevopoulou, Zoe and Pollack, Randy and Belanger, Olivier Savary and Sozeau, Matthieu and Weaver, Matthew},
  booktitle={The third international workshop on Coq for programming languages (CoqPL)},
  year={2017}
}

@article{paulin2015introduction,
  title={Introduction to the calculus of inductive constructions},
  author={Paulin-Mohring, Christine},
  journal={All about Proofs, Proofs for All},
  volume={55},
  year={2015},
  publisher={College Publications}
}

@online{bloomberg_crane_maintainers,
  author       = {Joomy Korkut and Matthew Weaver},
  title        = {crane},
  year         = {2026},
  url          = {https://github.com/bloomberg/crane#maintainers},
}

@article{10.1145/3656379,
author = {Forster, Yannick and Sozeau, Matthieu and Tabareau, Nicolas},
title = {Verified Extraction from Coq to OCaml},
year = {2024},
issue_date = {June 2024},
publisher = {Association for Computing Machinery},
address = {New York, NY, USA},
volume = {8},
number = {PLDI},
url = {https://doi.org/10.1145/3656379},
doi = {10.1145/3656379},
month = jun,
articleno = {149},
numpages = {24}
}

@online{riscv_unpriv_rv64,
  author       = {{RISC-V International}},
  title        = {Unprivileged ISA: RV64},
  year         = {2026},
  url          = {https://docs.riscv.org/reference/isa/unpriv/rv64.html},
  urldate      = {2026-04-28},
  organization = {RISC-V International},
  note         = {RISC-V ISA documentation (online reference)}
}

@article{bartholomew2006qemu,
  title={Qemu: a multihost, multitarget emulator},
  author={Bartholomew, Daniel},
  journal={Linux Journal},
  volume={2006},
  number={145},
  pages={3},
  year={2006},
  publisher={Belltown Media Houston, TX}
}

@article{quynh2015unicorn,
  title={Unicorn: Next generation cpu emulator framework},
  author={Quynh, NGUYEN Anh and Vu, DANG Hoang},
  journal={BlackHat USA},
  volume={476},
  year={2015}
}

@software{formal_land_coq_of_ocaml,
  author       = {{formal-land}},
  title        = {coq-of-ocaml},
  year         = {2026},
  url          = {https://github.com/formal-land/coq-of-ocaml},
}

@software{formal_land_rocq_of_rust,
  author       = {{formal-land}},
  title        = {rocq-of-rust},
  year         = {2026},
  url          = {https://github.com/formal-land/rocq-of-rust},
}

@misc{beg2026evaluatingllmgeneratedacslannotations,
      title={Evaluating LLM-Generated ACSL Annotations for Formal Verification}, 
      author={Arshad Beg and Diarmuid O'Donoghue and Rosemary Monahan},
      year={2026},
      eprint={2602.13851},
      archivePrefix={arXiv},
      primaryClass={cs.SE},
      url={https://arxiv.org/abs/2602.13851}, 
}

@article{DBLP:journals/pacmpl/MundlerHWSSV25,
  author       = {Niels M{\"{u}}ndler and
                  Jingxuan He and
                  Hao Wang and
                  Koushik Sen and
                  Dawn Song and
                  Martin T. Vechev},
  title        = {Type-Constrained Code Generation with Language Models},
  journal      = {Proc. {ACM} Program. Lang.},
  volume       = {9},
  number       = {{PLDI}},
  pages        = {601--626},
  year         = {2025},
  url          = {https://doi.org/10.1145/3729274},
  doi          = {10.1145/3729274},
  bibsource    = {dblp computer science bibliography, https://dblp.org}
}

@article{10.1145/3371119,
author = {Xia, Li-yao and Zakowski, Yannick and He, Paul and Hur, Chung-Kil and Malecha, Gregory and Pierce, Benjamin C. and Zdancewic, Steve},
title = {Interaction trees: representing recursive and impure programs in Coq},
year = {2019},
issue_date = {January 2020},
publisher = {Association for Computing Machinery},
address = {New York, NY, USA},
volume = {4},
number = {POPL},
url = {https://doi.org/10.1145/3371119},
doi = {10.1145/3371119},
journal = {Proc. ACM Program. Lang.},
month = dec,
articleno = {51},
numpages = {32}
}

@inproceedings{DBLP:conf/lpar/Leino10,
  author       = {K. Rustan M. Leino},
  editor       = {Edmund M. Clarke and
                  Andrei Voronkov},
  title        = {Dafny: An Automatic Program Verifier for Functional Correctness},
  booktitle    = {Logic for Programming, Artificial Intelligence, and Reasoning - 16th
                  International Conference, LPAR-16, Dakar, Senegal, April 25-May 1,
                  2010, Revised Selected Papers},
  series       = {Lecture Notes in Computer Science},
  pages        = {348--370},
  publisher    = {Springer},
  year         = {2010},
  url          = {https://doi.org/10.1007/978-3-642-17511-4\_20},
  doi          = {10.1007/978-3-642-17511-4\_20},
  bibsource    = {dblp computer science bibliography, https://dblp.org}
}

@inproceedings {AFLplusplus-Woot20,
author = {Andrea Fioraldi and Dominik Maier and Heiko Ei{\ss}feldt and Marc Heuse},
title = {{AFL++}: Combining Incremental Steps of Fuzzing Research},
booktitle = {14th {USENIX} Workshop on Offensive Technologies ({WOOT} 20)},
year = {2020},
publisher = {{USENIX} Association},
month = aug,
}

@inproceedings{DBLP:conf/nips/YangSGCSYGPA23,
  author       = {Kaiyu Yang and
                  Aidan M. Swope and
                  Alex Gu and
                  Rahul Chalamala and
                  Peiyang Song and
                  Shixing Yu and
                  Saad Godil and
                  Ryan J. Prenger and
                  Animashree Anandkumar},
  editor       = {Alice Oh and
                  Tristan Naumann and
                  Amir Globerson and
                  Kate Saenko and
                  Moritz Hardt and
                  Sergey Levine},
  title        = {LeanDojo: Theorem Proving with Retrieval-Augmented Language Models},
  booktitle    = {Advances in Neural Information Processing Systems 36: Annual Conference
                  on Neural Information Processing Systems 2023, NeurIPS 2023, New Orleans,
                  LA, USA, December 10 - 16, 2023},
  year         = {2023},
  url          = {http://papers.nips.cc/paper\_files/paper/2023/hash/4441469427094f8873d0fecb0c4e1cee-Abstract-Datasets\_and\_Benchmarks.html},
  bibsource    = {dblp computer science bibliography, https://dblp.org}
}

@inproceedings{
    jimenez2024swebench,
    title={{SWE}-bench: Can Language Models Resolve Real-world Github Issues?},
    author={Carlos E Jimenez and John Yang and Alexander Wettig and Shunyu Yao and Kexin Pei and Ofir Press and Karthik R Narasimhan},
    booktitle={The Twelfth International Conference on Learning Representations},
    year={2024},
    url={https://openreview.net/forum?id=VTF8yNQM66}
}

@article{achiam2023gpt,
  title={Gpt-4 technical report},
  author={Achiam, Josh and Adler, Steven and Agarwal, Sandhini and Ahmad, Lama and Akkaya, Ilge and Aleman, Florencia Leoni and Almeida, Diogo and Altenschmidt, Janko and Altman, Sam and Anadkat, Shyamal and others},
  journal={arXiv preprint arXiv:2303.08774},
  year={2023}
}

@misc{research2026composer2technicalreport,
      title={Composer 2 Technical Report}, 
      author={Cursor Research and : and Aaron Chan and Ahmed Shalaby and Alexander Wettig and Aman Sanger and Andrew Zhai and Anurag Ajay and Ashvin Nair and Charlie Snell and Chen Lu and Chen Shen and Emily Jia and Federico Cassano and Hanpeng Liu and Haoyu Chen and Henry Wildermuth and Jacob Jackson and Janet Li and Jediah Katz and Jiajun Yao and Joey Hejna and Josh Warner and Julius Vering and Kevin Frans and Lee Danilek and Less Wright and Lujing Cen and Luke Melas-Kyriazi and Michael Truell and Michiel de Jong and Naman Jain and Nate Schmidt and Nathan Wang and Niklas Muennighoff and Oleg Rybkin and Paul Loh and Phillip Kravtsov and Rishabh Yadav and Sahil Shah and Sam Kottler and Alexander M Rush and Shengtong Zhang and Shomil Jain and Sriram Sankar and Stefan Heule and Stuart H. Sul and Sualeh Asif and Victor Rong and Wanqi Zhu and William Lin and Yuchen Wu and Yuri Volkov and Yury Zemlyanskiy and Zack Holbrook and Zhiyuan Zhang},
      year={2026},
      eprint={2603.24477},
      archivePrefix={arXiv},
      primaryClass={cs.SE},
      url={https://arxiv.org/abs/2603.24477}, 
}

@article{lattuada2023verus,
  title={Verus: Verifying rust programs using linear ghost types},
  author={Lattuada, Andrea and Hance, Travis and Cho, Chanhee and Brun, Matthias and Subasinghe, Isitha and Zhou, Yi and Howell, Jon and Parno, Bryan and Hawblitzel, Chris},
  journal={Proceedings of the ACM on Programming Languages},
  volume={7},
  number={OOPSLA1},
  pages={286--315},
  year={2023},
  publisher={ACM New York, NY, USA}
}


\end{document}